\documentstyle[pre,aps,twocolumn]{revtex}
\begin{document}
\draft
\title{Crossover from Percolation to Self-Organized Criticality}
\author{Barbara Drossel, Siegfried Clar, and Franz Schwabl}
\address{Institut f\"ur Theoretische Physik, \\
         Physik-Department der Technischen Universit\"at M\"unchen, \\
         James-Franck-Str., D-85747 Garching, Germany}
\date{\today}
\maketitle
\begin{abstract}
We include immunity against fire as a new parameter into the  self-organized 
critical forest-fire model.  When the immunity assumes a critical value,
clusters of burnt trees are identical to percolation clusters of random bond
percolation. As long as the immunity is below its critical value, the
asymptotic critical exponents are those of the original self-organized critical
model, i.e. the system performs a crossover from percolation to self-organized
criticality. We present a scaling theory and computer simulation results.
\end{abstract}
\pacs{PACS numbers: 05.40.+j, 05.70.Jk, 05.70.Ln}

%%%%%%%%%%%%%%%%%%%%%%%%%%%%%%%%%%%%%%%%

\narrowtext

Several extended nonequilibrium systems show 
scaling behavior over a wide range of parameter values and independently of the 
initial conditions. They are called {\it self-organized critical} (SOC) 
\cite{bak1}
and might explain the ubiquity of fractal structures in nature. By analogy to 
equilibrium critical phenomena, the question arises if SOC phenomena are
universal, i.e. if the critical behavior depends only 
on few properties of the model as dimension, conservation laws, or number of 
components. So far, examples for both  universal and nonuniversal behavior have
been found. While the critical behavior of the sandpile model \cite{bak1}
seems to be robust with respect to various changes, the critical exponents of 
an earthquake model \cite{chr1} change continuously as function of a parameter
which characterizes the degree of conservation. In equilibrium critical 
phenomena, such a continuous change of critical exponents occurs only in
exceptional cases, e.g. the eight vertex model.
Instead, critical behavior is characterized by a crossover between 
two fixed points, when certain parameters are changed. This crossover is
described by scaling functions and a crossover  
exponent $\phi$. In this letter, we report for the first time a crossover 
phenomenon in SOC systems by including a new parameter, the immunity $g$, 
in the SOC forest-fire model \cite{dro1}. $g$ is the probability that fire 
cannot pass from one tree to a neighboring tree. When the immunity assumes a
critical value $g_c$, the forest becomes very dense, and clusters that are
burnt by a lightning stroke are identical to percolation clusters. 
As long as the immunity is below its critical value, the asymptotic critical
exponents of the model are those of the original SOC
forest-fire model, i.e. the system performs a crossover from percolation to
SOC.

The forest-fire model is defined on a $d$-dimensional hypercubic lattice with 
$L^d$ sites. Each site is either occupied by a tree, a burning tree, or it is 
empty. The state of the system is parallely updated according to the following 
rules:
\begin{enumerate}
\item Burning tree $\longrightarrow$ empty site.
\item Tree $\longrightarrow$ burning tree with probability $1-g^n$ if $n \ge 
1$ nearest neighbors are burning. 
\item Tree $\longrightarrow $ burning tree with probability $f$ if no nearest 
neighbor is burning.
\item Empty site $\longrightarrow$ tree with probability $p$.
\end{enumerate}
Starting with arbitrary initial conditions,  after some time the system
approaches a steady state the properties of which depend on the parameter
values but not on the initial state. We always chose the lattice size so large
that the steady state is also independent of the boundary conditions. 
Let $\rho_t$ be the density of trees, $\rho_e$ the density of empty sites, and 
$\rho_f$ the density of burning trees in the steady state. Large forests and 
therefore large fires occur only if the lightning probability $f$ is much 
smaller than the tree growth probability $p$. If additionally 
large and small fires look alike, the system becomes scale invariant. Small
fires live only for few time steps  
and are extinguished before new trees grow in their neighborhood. Scale 
invariance therefore can only be observed if tree growth is so slow that 
large fires also are extinguished before new trees grow at the edge of the 
burning forests. We conclude that SOC behavior occurs in the forest-fire model 
if the following conditions are satisfied: Lightning occurs seldom compared to
tree growth,  
and tree growth is much slower than the lifetime of a fire, i.e.
\begin{equation}
f\ll p \ll (f/p)^{\nu'} \label{eq1}
\end{equation}
with an appropriate exponent $\nu'$ \cite{dro1}.
Eq. (\ref{eq1}) represents a double separation of timescales.
Since in the steady state the mean number of burning trees equals the mean 
number of growing trees, the mean number of trees destroyed by a lightning 
stroke is \cite{dro1}
\begin{equation}
\bar s = {\rho_e \over \rho_t} {p\over f} \simeq {1-\rho_t \over \rho_t} 
{p\over f}\, . \label{eq2}
\end{equation}
In the last step, we have neglected the fire density $\rho_f$ which vanishes in
the limit of perfect time scale separation.
In $d\ge 2$ dimensions and for values of the immunity below its critical
value, the  critical forest density
 $\rho_t^c= \lim_{f/p \to 0} \rho_t$ is smaller than 1, and Eq. (\ref{eq2}) 
 represents a power law
$
\bar s \propto (f/p)^{-\gamma} $
with $\gamma=1$, indicating a critical point at $f/p=0$. 
Close to this critical point, i.e. if $f \ll p$, there is scaling over many
orders of magnitude. 

Several critical exponents characterizing this
scaling behavior have been defined \cite{dro1,hen1,cla1}. The forest density
satisfies for small $f/p$ a power law
\begin{equation}
\rho_t^c-\rho_t\propto (f/p)^{1/\delta}, \label{eq2a}
\end{equation}
as first stated in \cite{hen1} and \cite{gra1}.

 Let $s$ be the number of trees destroyed by a fire. The 
size distribution of fires is a power law \cite{dro1}
\begin{equation}
sn(s) \propto s^{1-\tau} {\cal C}(s/s_{\text{max}}) \label{eq3}
\end{equation}
with 
\begin{equation}
s_{\text{max}} \propto (f/p)^{-\lambda}. \label{eq4}
\end{equation}
Inserting Eq. (\ref{eq4}) and Eq. (\ref{eq3}) into Eq. (\ref{eq2}), we obtain 
the scaling relation
$\lambda = 1/(3-\tau)$,
which is valid as long as $\rho_t^c < 1$. In general, if we allow also the case
$\rho_t^c=1$, as relevant for $g=g_c$,  $\bar s$ is given by a power law
$\bar s \propto (f/p)^{-\gamma}$, with a value for $\gamma$ which may be
different from 1,  and we have the more general scaling relation 
\begin{equation}
\lambda= \gamma / (3-\tau)\, .  \label{eqsc1}
\end{equation}

We also define the radius $R(s)$ of a forest cluster that has just 
been burnt down. Its fractal dimension $\mu$ is given by
\begin{equation}
R(s) \propto s^{1/\mu}. \label{eq5}
\end{equation}
The  correlation length $\xi$ is defined by
\begin{equation}
\xi^2 = {2\sum_{s=1}^{\infty} R^2(s) s^2 n(s) \over \sum_{s=1}^\infty s^2n(s)}.
\label{eq6} 
\end{equation}    
Together with Eqs. (\ref{eq3}), (\ref{eq4}) and (\ref{eq5}), this gives 
\begin{equation}
\xi \propto (f/p)^{-\nu} \hbox{ with } \nu = \lambda/ \mu\, . \label{eqsc2}
\end{equation}

In the SOC forest-fire model without immunity, these exponents have already
been determined by computer  
simulations \cite{hen1,cla1,gra1,chr2}. In $d=2$ dimensions, they are
\cite{cla1} 
\begin{eqnarray}
\tau_{\text{soc}} &=& 2.14(3), \, \lambda_{\text{soc}}= 1.15(3),\,
\mu_{\text{soc}}=  
1.96(1), \nonumber \\ \nu_{\text{soc}}&=& 0.58,\,
1/\delta_{\text{soc}}=0.48(2). 
\label{exp1} 
\end{eqnarray}
In $d=1$ dimension, their values have been 
derived analytically \cite{dro2}. In \cite{cla1}, it is also shown that the 
critical exponents in two dimensions do not change when the lattice symmetry is
changed or next-nearest neighbor interaction is included. 

We now consider the SOC forest-fire model with nonvanishing immunity, i.e. for
$g>0$. Immunity was first  
introduced in \cite{dro3} into a version of the  forest-fire model \cite{bak2} 
which is not SOC \cite{gra2,mos1}. This model shows a   percolation-like phase
transition  
when the immunity approaches a critical value $g_c$. In the present paper, 
the immunity is for the first time included into the SOC forest-fire model. It
is defined differently than before: While  in 
\cite{dro3} the immunity is a property of trees,  in the present 
paper it is a property of bonds between neighboring trees (see rule 2. above). 
This has two advantages: The simulation program is less complicated, and the
critical immunity is known to be exactly $g_c = 0.5$  
which is just 1 minus the percolation threshold for bond percolation.

When the immunity is different from 0, not all trees that are neighbors of a 
burning tree catch fire, and consequently the fire does no longer burn forest
clusters but clusters of trees that are 
connected by non immune bonds. With increasing immunity, the 
forest density increases, since fewer trees are burnt. At the critical 
immunity $g_c$, the critical forest density is $\rho_t^c = 1$. Then we have the
following situation: 
The forest is completely dense in the limit $f/p \to 0$, and clusters that are
destroyed by fire are 
percolation clusters of bond percolation. Consequently the exponents $\tau$ and
$\mu$ are 
given by percolation theory 
\begin{equation}
\tau(g_c) \equiv \tau_c = \tau_{\text{perc}}= 187/91 \simeq 2.05 \label{eq7}
\end{equation}
and
\begin{equation}
\mu(g_c)\equiv \mu_c = \mu_{\text{perc}}= 91/48 \simeq  1.90. \label{eq8}
\end{equation}
Bonds that are immune during one time step might be non immune during the
next one, and consequently lightning strokes at the same site burn down
different clusters  at different times.
When $f /p$ is finite, there is a cutoff in cluster size, since large fires 
are stopped by empty sites that have been left from earlier fires. 
The mean forest density is no longer 1. We determined the critical 
exponents $\lambda$, $\delta$, and $\nu$  at $g=g_c$ by computer simulations in
$d=2$ dimensions
and obtained 
\begin{equation}
\lambda_c = 0.92(3),\, 1/\delta_c = 0.15(1) ,\, \nu_c = 0.484(2). \label{eq9}
\end{equation}
The scaling relations Eq. (\ref{eqsc1}) and Eq. (\ref{eqsc2}) still hold at
$g=g_c$ and are confirmed by our simulations.
Using Eqs. (\ref{eq2}) and (\ref{eq2a}), we obtain
\begin{equation}
\bar s \propto (f/p)^{-\gamma_c} \text{ with } \gamma_c = 1 - 1/\delta_c \, .
\label{eqgammac}
\end{equation}
Our simulations yield $\gamma_c = 0.84(2)$, in agreement with Eq.
(\ref{eqgammac}). 

When the immunity is just below its critical 
value ($(g_c-g) \ll 1$), the situation becomes more complicated. On small
length scales, a system close to the percolation threshold  
cannot be distinguished from a system exactly at the percolation threshold. On 
large length scales, however, the difference can be seen. If the initial state 
were a completely dense forest, a finite fraction of all trees were connected 
by non immune bonds for $g<g_c$. This infinite cluster would soon be destroyed 
by lightning and would never occur again due to Eq. (\ref{eq2}). Consequently
the critical forest density is $\rho_t^c < 1$ for $g<g_c$. Imagine for a moment
that bonds are permanently immune or   
non immune (in reality, rule 2. says that they might be immune during one time 
step and non immune during the next one). Then the dynamics on the sites which 
initially belonged to the infinite cluster are completely decoupled from 
dynamics on the remaining sites. We can consider the sites which belonged to 
the infinite cluster as an  independent subsystem where no immunity exists. 
They form a two-dimensional lattice with another symmetry than the original
hypercubic one. Since  
we already know from earlier simulations that the exponents of  the SOC 
forest-fire model do not change when lattice symmetry is changed, we expect 
that the critical exponents on our subsystem are just the ones which we 
obtained for $g=0$ and which have been assigned an index "soc".  The remaining
sites are connected by non immune bonds to form finite clusters of bond
percolation. On each of these clusters, the dynamics are independent of the
other clusters and of the infinite subsystem. The largest of these clusters
have a radius of the order of the percolation correlation length
$\xi_{\text{perc}} \propto (g_c-g)^{-\nu_{\text{perc}}}$. When $f/p$ becomes
very small, these finite clusters are very rarely struck by lightning, and
consequently the mean tree density on these clusters approaches the value 1 in
the 
limit $f/p \to 0$. Each time lightning strikes such a cluster, there is a fire
which has the size and the fractal dimension of a finite percolation cluster. 
As long as $f/p$ is so small that 
the correlation length $\xi \propto (f/p)^{-\nu_c}$ is much larger than the 
percolation correlation length $\xi_{\text{perc}}$, all large fires occur on
the infinite subsystem, and the  
exponents $\nu$, $\delta$, and $\lambda$ are those of the subsystem, i.e. 
the SOC ones. The exponents $\tau$ and $\mu$ of the large 
fires are by the same reasoning also the SOC ones. But for fires with a 
radius smaller than the percolation correlation length, $\tau$ and $\mu$ assume
their  percolation values since the infinite subsystem contains only a small
portion of all sites and consequently most of the small fires occur on the
finite clusters.  When $f/p$ becomes so large that the  forest clusters on
our infinite subsystem are no more larger than the percolation correlation
length, 
our system is dominated by the dynamics on the finite clusters 
and becomes indistinguishable from a system at $g=g_c$.  
The exponents 
$\nu$, $\delta$, and $\lambda$ then are identical to those at $g=g_c$, and the 
exponents $\tau$ and $\mu$ are those of percolation theory on all length 
scales up to the correlation length.
So far, we considered the case of small $g_c-g$. When the immunity is far away 
from its critical value, the infinite subsystem contains a large portion of 
all sites, and the percolation-dominated behavior cannot occur any more. 

Unfortunately, all these considerations are based on the assumption that bonds 
are permanently immune or non immune which  in reality is not the case. 
Consequently, there exists no subsystem which is decoupled from the rest of 
the system.  Nevertheless, the main conclusions should remain valid: On 
length  scales smaller than the percolation correlation length, the system
cannot be distinguished from a system at $g=g_c$. When $f/p$ becomes very
small, there are fires which spread further 
 than the percolation correlation length. These fires are stopped by empty 
sites that were created by earlier fires. This is again the same mechanism as 
in the limit $g=0$ or on the infinite subsystem in the case of fixed immune
bonds:  
fires that would spread indefinitely  if there were no empty sites are stopped 
by empty sites. We conclude that these large fires lead again to the critical 
exponents $\lambda_{\text{soc}}$, $\nu_{\text{soc}}$, and 
$\delta_{\text{soc}}$.

Led by these considerations, we  make the following scaling ansatz for the
correlation length:
\begin{equation}
\xi = (f/p)^{-\nu_c} F\left({g_c-g \over (f/p)^\phi}\right).
\label{eq11}
\end{equation}
It is plausible that the crossover from percolation-like to SOC behavior takes
place when  
$f/p$ becomes so small that the correlation length exceeds the percolation 
correlation length, which suggests that the crossover exponent $\phi$ is 
\begin{equation}
\phi= \nu_{c}/\nu_{perc}\, . \label{eqsc3}
\end{equation}
The scaling function $F(x)$ is constant for small $x$ and is $\propto 
x^{(\nu_{soc}-\nu_c)/\phi}$ for large $x$. 
Analogous scaling laws hold for $s_{\text{max}}$ and $\rho_t^c-\rho_t$.
We already mentioned above that the critical forest density is $\rho_t^c=1 $ at
$g_c$.  
We therefore expect an additional power law 
\begin{equation}
1-\rho_t^c(g) \propto (g_c-g)^y\, . \label{eq12}
\end{equation}
The exponent $y$ is obtained from the scaling ansatz
\begin{equation}
1-\rho_t = (f/p)^{1/\delta_c} G\left({ g_c-g \over (f/p)^\phi}\right) . 
\label{eq13}
\end{equation}
In the limit $f/p\to 0$, the forest density becomes independent of $f/p$ and
assumes a value $\rho_t^c \neq 1$. Therefore 
$G(x) \propto x^{1/\phi\delta_c}$ for large $x$, yielding
\begin{equation}
y = \nu_{\text{perc}} / \nu_c\delta_c \, . \label{eqsc4}
\end{equation}

Our simulations confirm all these results. They were performed using the same
method as in \cite{cla1,gra1} for lattices of up to $8192^2$ sites and values
of $f/p$ down to $10^{-6}$. The immunity varied between $(g_c-g)=0.02$ and
$(g_c-g)=0.001$. The range of $f/p$ is limited by
finite-size effects for small $f/p$ and noncritical behavior for large $f/p$,
the range of $g$ was selected from the condition that at least part of the
crossover region is covered.
Fig. \ref{fig1} shows the scaling function for the correlation length $F(x)$
for  
different values of $g_c-g$. The scaling ansatz Eq. (\ref{eq11}) is well 
confirmed since all curves coincide. The dashed line represents $F(0)$ as
obtained from the simulations at $g_c$. 
We also checked the scaling relation Eq. (\ref{eqsc4}). 
 Fig. \ref{fig2} shows
the critical forest density as function of $g_c$. We obtain $y = 0.43(2)$, in
agreement with Eq. (\ref{eqsc4}) (remember that $\nu_{\text{perc}} = 4/3$).
Since the difference between $\tau_{\text{soc}}$ and $\tau_{\text{perc}}$, as
well as between $\mu_{\text{soc}}$ and $\mu_{\text{perc}}$ is very small, the
crossover behavior of $n(s)$ and $R(s)$ could not be evaluated.

In \cite{cla1}, we also defined  exponents $\nu'$, $\mu'$ and $\alpha$
describing the temporal behavior of the fires. The crossover in $\mu'$ and
$\nu'$, which enters the condition for time scale separation Eq. (\ref{eq1}),
is analogous to the crossover in $\mu$ and $\nu$ \cite{footnote}. The change in
$\alpha$  is too small for  any crossover to be observable.
 
To conclude, we have shown by analytic arguments and by computer simulations
that the 
forest-fire model performs a crossover from percolation to SOC when the
immunity is close to its critical value. This crossover is characterized by
scaling 
functions which are defined in the same way as in crossover phenomena at
equilibrium phase transitions.

Although all simulations were performed in $d=2$ dimensions, we expect that 
this crossover behavior can also be observed in higher dimensions. In $d=1$, 
the critical immunity is $g_c=0$, and no crossover can take place. For 
$d\ge 6$, simulations suggest that the critical exponents assume their
mean-field values which are identical to those of percolation 
\cite{cla1,chr2}. Consequently there is no crossover in $d \ge 6$ dimensions.

\acknowledgements
We thank E. Frey for useful discussions.

\begin{figure}
\caption{Crossover scaling function $F(x)$ for the correlation length for
different 
values of the immunity. The dashed line represents $F(0)$ as obtained at
$g=g_c$.} 
\label{fig1}
\end{figure}

\begin{figure}
\caption{The critical forest density as function of the immunity.}
\label{fig2}
\end{figure}


\begin{references}
\bibitem{footnote} The values of these exponents are in $d=2$ dimensions
$\nu'_{\text{soc}}=0.61$, $\mu'_{\text{soc}}=1.89(3)$, $\nu'_c=0.55$, and
$\mu'_c= 1.68$.
\end{references}
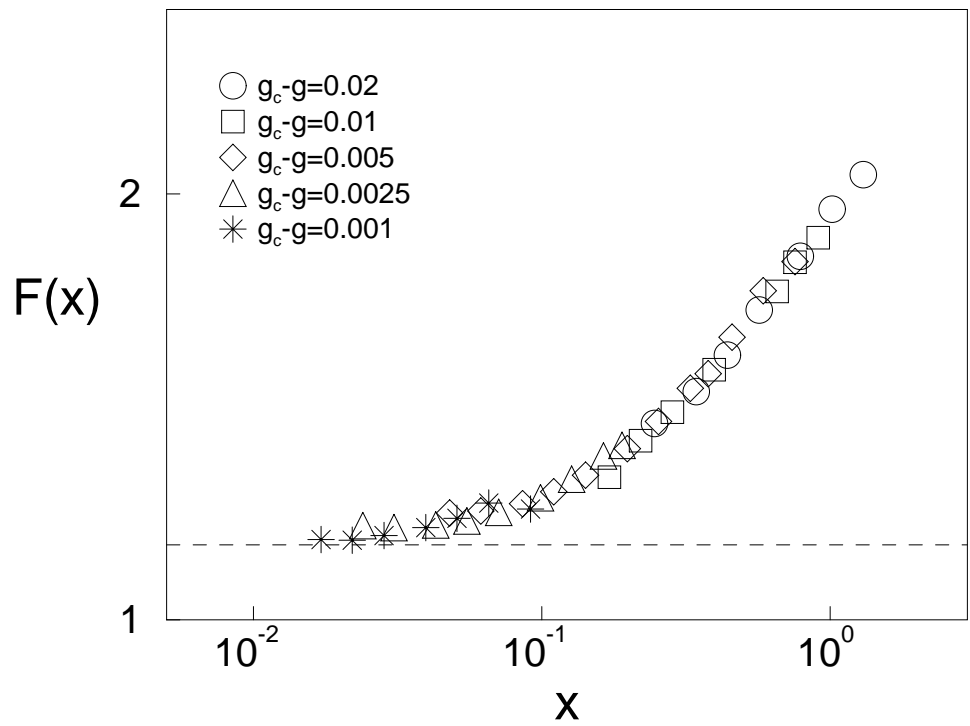
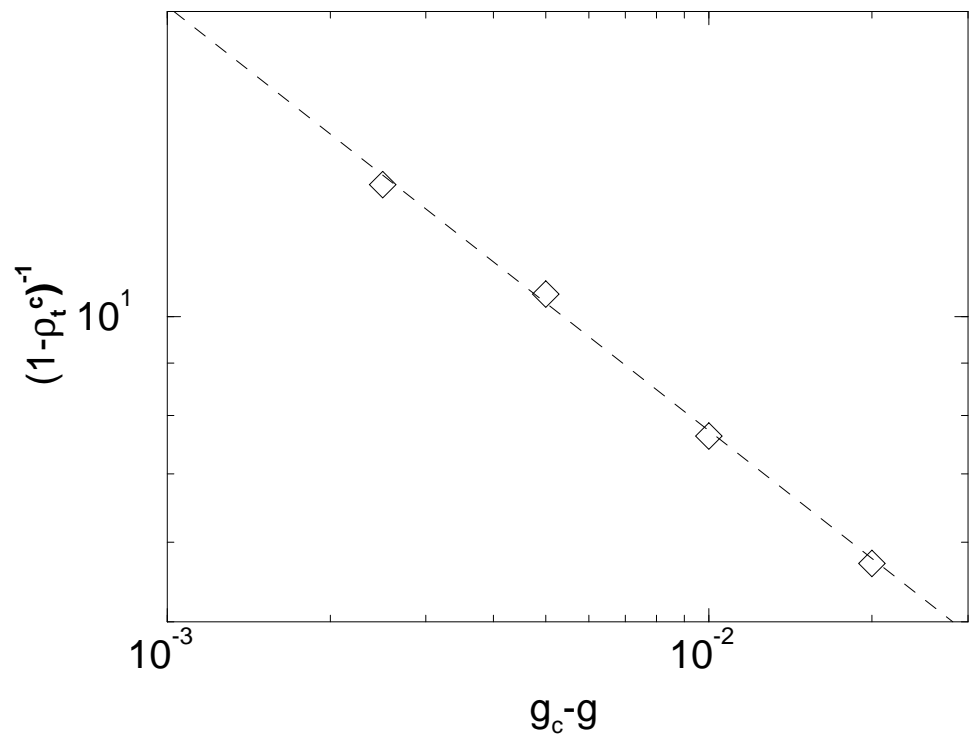
\end{document}